# Trust and Privacy in Knowledge Graphs[1]

Daniel Schwabe

    Department of Informatics, Pontifical Catholic University, Rio de Janeiro, Brazil
    dschwabe@inf.puc-rio.br

Carlos Laufer

    Department of Informatics, Pontifical Catholic University, Rio de Janeiro, Brazil
    laufer@globo.com

**ABSTRACT**

This paper presents the KG Usage framework, which allows the introduction of KG features to support Trust, Privacy and Transparency concerns regarding the use of its contents by applications. A real-world example is presented and used to illustrate how the framework can be used.

## 1 Introduction

The term "Knowledge Graph" (KG) was introduced by Google in 2012[2], although to this day there isn't a precise definition of the term [11] Nevertheless, graph-based databases were available before this (e.g., Wordnet [23] DBPedia [19] Yago [34], CYC [20] NELL [7] and additional ones (e.g. ConceptNet [32]) continue to be created on a regular basis.

While the graph model or some variant has been used in several KGs, it has already been observed that using only nodes as the "granule" of information is too low to express complex types of information, such as events, or time-varying data. For example, Wikidata [38] is organized around Items described by a collection of Statements [12] Another reason for having more complex "granules" is recording provenance (meta) data, which is a fundamental part of data in some domains such as life-sciences [17].

KGs differ also on the way they are built (populated). A few are curated (e.g., CYC), others rely on crowdsourced information (e.g. Wikidata); most extract information from structured, semi-structured or textual information harvested from the Web.

The multiplicity of sources and various extraction approaches naturally raises the issue of data quality and confronts the user of the data in the KG with the issue of trusting, or not, the information obtained from the KG. For some types of information, for example in case of online reviews online and social media, this trust

---

can have a direct effect on commercial success (e.g. [2]). This highlights the fact that data, ultimately, expresses a belief, opinion or point of view of some agent.

From a broader perspective, information (and knowledge) has become the prime resource in the Third Industrial Revolution, also called the Digital Age – when digital technologies enabled new ways of generating, processing and sharing information [9][26], and is becoming more so as we move into the Fourth Industrial Revolution (4IR)[28]. The 4IR is characterized by a fusion of technologies, which is blurring the lines between the physical, digital, and biological spheres.

Increasingly, systems and applications operate in a context which the flow of information has direct bearing on daily lives of billions of people, where two fundamental characteristics of the use of such information emerge – Transparency and Privacy.

Transparency can be seen as the quality that allows participants of a community to know what the particular processes and agents are that are being used in its functioning. It is generally regarded as a means to enable checks and balances within this community, ultimately providing a basis for trust among participants. Considering the community as the whole society, these checks and balances are reflected in its political system, to prevent misuse by any of the parties involved.

One of the mechanisms created to increase transparency in political systems is the enactment of legislation ensuring the right of its members to access to information in a variety of contexts, ranging from government-produced information and data to consumer-related information regarding goods and products, as well as the right of individuals to freely create, publish and access information.

The free flow of information, on the other hand, may conflict with another basic human right, that of Privacy [37] There are many definitions for Privacy [24] , but in essence they all refer to the right of an individual to control how information about her/him is used by others.

In order to deal with the myriad of often conflicting cross-cutting concerns, Internet applications and systems must incorporate adequate mechanisms to ensure compliance of both ethical and legal principles.

In order to be effective, we claim that the *use* of Knowledge Graphs must support these concerns – trust, privacy and transparency. In this paper we propose a framework that enables this support.

## 2 Background Concepts

Before detailing our framework, we briefly present our definition for each of the basic concerns. We have detailed each of them in other publications, as referenced in each sub-section.

## 2.1 Trust

The issue of trust has been prevalent in the Internet since its popularization in the early 90s (see [14] for a survey), with a focus on the lower layers of the Internet Architecture, emphasizing authentication. More recently, with the advent of the Web and social networks, the cybersphere, and society as a whole, has become heavily influenced by information (and misinformation) that flows in news sites and social networks in the Internet. There are many studies carried out in several disciplines attempting to characterize and understand the spread of information in the cybersphere, as well as how this affects society (see [21]for an overview).

The original vision for the Semantic Web included a "Trust" layer, although its emphasis was more on authentication and validation with static trust measures for data. There have been many efforts in representing trust, including computational models - a general survey can be found in [25]; [3]presents an excellent earlier survey for the Semantic Web; and [30] surveys trust in social networks. In the Linked Data world, it is clear that facts in Semantic Web should be regarded as claims rather than hard facts (e.g., [6]), which naturally raises the issue of trust on those claims.

As proposed initially in [1] and later in [18] and [29] the approach used is based on the work of Gerck [13] and Castelfranchi et al. [8], taking the view that trust is "knowledge-based reliance on received information", that is, an agent decides to trust (or not) based solely on her/his knowledge, and the decision to trust implies the decision to rely on the truth of received or on already known information to *perform some action*.

In terms of a Knowledge Graph, an agent wishing to perform an action must first filter those information items it deems "trusted", i.e., it will use them to perform the intended action. Since it is not possible to "half-act", in this sense trust is binary – either the agents trusts the information, or it does not. A more extensive discussion can be found in **Error! Reference source not found.**.

Trusted information is the basis for supporting privacy and transparency, as discussed next.

## 2.2 Privacy

There are many definitions for privacy - see for example [31]; [24][33] present surveys on multi-party privacy. Tavani [35] classified privacy theories into four broad types: the nonintrusion, seclusion, limitation, and control theories. For our research, we adopt the Restricted Access/Limited Control (RALC) Theory proposed by Moor and Tavani [36]. RALC presupposes that an adequate theory of privacy needs to differentiate the concept of privacy from both the justification and the management of privacy. In this sense, the RALC framework has three components: an account of the concept of privacy, an account of the justification of privacy, and an account of the management of privacy. "*RALC requires that one must have protection from intrusion and interference and information access, it addresses concerns not only about protecting informational privacy (as described in the control and the limitation theories) but also about protection against the kinds of threats described in the nonintrusion and the seclusion theories as well.*" [35]. Based on this theory, we define Privacy as *"controlled access to information related to an agent"*. In order to ensure privacy, it is necessary to answer three questions:

- Q1: What types of Actions are allowed (and controlled) over Knowledge Items (KIs)?
- Q2: What are the relation types, between some Agent and an KI, which entitle this Agent to establish a Privacy Rule governing Actions over that KI?
- Q3: How to resolve conflicts between applicable rules?

## 2.3 Transparency

Generally speaking, according to Meijer [22] and others [27], transparency can be defined as "the availability of information about an actor that allows other actors to monitor the workings or performance of the first actor." It contemplates the "capacity of outsiders to obtain valid and timely information about the activities of government or private organizations"[3].

Transparency presupposes the involvement of an observed and an observer [5]. In contrast to privacy, that is concerned with information about individuals, transparency concerns any type of information, although it may make a difference if the producer (author) is an organization or an individual [16].

Transparency is also related to privacy. In the so-called "Attention Economy" [10], for example, the information about users and consumers are a primary source of value, and companies actively seek to obtain as much information about users. This can be in direct conflict with privacy rights of users, who have the right to control how information about him/her is used by others. In this state of affairs, transparency can support the resolution of potential conflict with privacy policies, since the disclosure of information about the company's processes and procedures associated to this individual contributes to trusting that it is indeed compliant with these regulations and thus stimulating authorization on the part of the users.

Seen in a more abstract manner, both Privacy and Transparency relate to controlling actions over information, and who can define such controls. As such, they can be regarded as two points in the same control dimension.

---

[3] https://www.britannica.com/topic/transparency-government#accordion-article-history

Privacy tends to limit or restrict actions over information items, whereas Transparency tends to allow (in some cases, mandate) actions over them, which explains the natural tension the exists between the two.

## 3 A Motivational Example - Disaster Relief Donation

The goal of this example is to illustrate the complex and interdependent nature trust, privacy and transparency; it will later be used to show the expressive power of the framework and how it incorporates the various concerns at play.

Consider a scenario in which a disaster has occurred, and Ed wants to donate some money to help with the disaster relief actions. Ed has received several donation requests from different organizations, therefore he must choose one to make the donation. However, given his past experience, Ed wants to make sure that the donation money will actually be used for the relief actions, rather than being misused, e.g., funding the organization's basic infrastructure, or employed in another action, or even pocketed by unscrupulous officers of the receiving organization.

Ed formulates a rule that says that he only trusts organizations that openly publish who are their financial officers, and their financial records. Financial records must be validated by accredited audit organizations.

Furthermore, because of personal reasons, Ed does not want to contribute to an organization in which George is an officer. This rule can be regarded as an application-related rule, akin to what is referred to as a "business rule" in traditional software development.

Let us assume that there is a law that stipulates that not-for-profit organizations must publicly identify their officers and that George is an officer of ReliefOrg, an NGO dedicated to raising funds for and helping disaster relief efforts. George, being a very reserved person, has a privacy rule that stipulates that his association with any organization, including ReliefOrg, should not be made public.

Ed received a request for donation from ReliefOrg and needs to decide whether he should donate or not.

The first step Ed follows is to verify is the officers of ReliefOrg are published in the KG[4]. Here we can see a potential conflict between transparency and privacy rules. George's relation with ReliefOrg should be accessible in the KG, according to not-for-profit legislation. On the other hand, George's privacy rules would prevent this access. Since transparency legislation in this case has higher precedence than personal rules, this association can be accessed by Ed's rules, and therefore ReliefOrg would not be accepted as a recipient of donations by Ed.

Consider now a slightly different scenario. Ed has no personal objections to George, but wants to check the financial integrity of ReliefOrg. He retrieves the financial report for ReliefOrg from the KG, but wants to make sure it has been audited, so he looks for a certification of the financial report. He finds out it has been audited by AuditInc, who he has not heard of, leading him to verify that it has an accreditation certificate from a public authority. If such a certificate is available, Ed analyses the financial report finding nothing wrong in principle.

However, tipped by a friend, Ed learns that George may in fact be one of the owners of AuditInc. He then checks AuditInc to see if its owners are listed, and whether George is one of them. Given George's privacy rule plus the fact that AuditInc is not a not-for-profit organization, his privacy rule would prevent access to the owner relationship, so Ed would not see that George is one of the owners, thus deciding to contribute. Note that in this scenario, George's privacy rule would apply not only to AuditInc's information, but also to information furnished by others, for example, a photo on a social network where George appears in the annual

---

[4] See for instance https://permid.org or https://opencorporates.com as examples of KGs with this type of information.

Christmas party of AuditInc with a caption mentioning his role as one of the partners, or perhaps a badge on his neck identifying him as such.

We next describe our proposed KG Usage framework and subsequently analyze this example showing how it can be represented by it.

## 4 A Summary of the KG Usage Framework.

Figure 2 shows a diagram of the use of information within a KG. Using a KG is represented by a Request made by some Agent for an Action over a Knowledge Item (KI).

We assume the existence of an underlying RDF graph, which would be equivalent to the "traditional" definition of KGs. The actual KG is formed by defining several named graphs over this underlying RDF graphs as a way to structure the RDF triples into Knowledge Items, similarly to Items and Qualifiers in Wikidata [12].

Thus, a Knowledge Graph represents a collection of interlinked descriptions of KIs – real-world objects, events, situations or abstract concepts – where:
- Descriptions have a formal structure that allows both people and computers to process them in an efficient and unambiguous manner;
- Entity descriptions contribute to one another, forming a network, where each entity represents part of the description of the entities, related to it.[5]

We propose to represent the KG as a collection of Knowledge Items (KIs), each of which as a nanopublication[6] [15].

## 4.1 KI Representation

A KI, as any nanopublication, comprises an *assertion graph*, a *provenance graph* and a *publication info* graph.

The *assertion* graph of a KI contains a set of assertions about its content. The assertions in this graph are a subset of the assertions in the underlying RDF graph. As an example, if a KI refers to an event, it would have statements about the participants and their roles, location information, date information, depictions (photos, videos, ...) and so on.

The *provenance* graph of the nanopublication will contain provenance information about the assertions in the assertion graph (e.g.; what image or natural language processing software was used, recorded location info, whether the assertions were inferred using some inference engine, etc.). The provenance graph can be used to represent, to the desired level of detail, the supporting information for the assertions. For example, if an automated face recognition algorithm was used, the provenance information represented in the *provenance* graph of the nanopublication. may inform which algorithm, which parameters were used in this particular case, and a confidence factor. Another use of provenance can be seen in the case of a statement stating that, for example, <Barack Obama> placeOfBirth <Hawaii>. The provenance information can include documentation to support its truthfulness, such as a reference to a birth certificate that states that indeed the place of birth of Barack Obama is Hawaii.

The *publication info* graph will contain metadata about the creation of the KI itself (as opposed to the information contained in its *assertions* sub-graph).

---

5 https://www.ontotext.com/knowledgehub/fundamentals/what-is-a-knowledge-graph/

6 http://nanopub.org

## 4.2 Controlling usage

Since both Privacy and Transparency refer to actions over some information (in a KI), an Authorization must be granted for this Request, according to the Rules set forth by stakeholders. Stakeholders include persons "related" to the KI, as well as institutional agents such as "the State" (whose rules are stated as laws). Rules may be based (drawn) on any information available in the KG. We refer to both Privacy and Transparency rules generically as Usage Rules.

Figure 1 present the basic algorithm to evaluate a Request. The first step in granting an Authorization for the action is to determine which KIs in the KG are trusted by, the Agent requesting the Action. This is realized by a Trust Engine that collects applicable rules and includes them in a Trusted sub-graph of the KG. We refer to the set of trusted KIs by Agent $a$ as the Trusted KG(a).

Once the Trusted KG has been determined, a set of usage rules are evaluated, which entails determining applicable rules (UsageRuleSet), evaluating them (EvalUsageRuleSet) and resolving conflicts if they arise. In order to evaluate an usage rule, it is also necessary to evaluate the trust rules of the author of that usage rule, which needs to be based on trusted KIs. Thus, a set of Trusted KG($a_i$) is defined, one for each Author $a_i$ of an applicable usage rule. The result of this process defines the final Authorization for the Action in the Request that was made. If the Authorization is "Allowed", the Action is carried out using the TrustedKG of the Agent that requested the Action.

Define EvalRequest(Agent, Action, KI),
1. Let TS <- TrustRuleSet(Agent), RS <- UsageRuleSet(KI).
2. Let TGA <- EvalTrustRuleSet (TS, Agent, Action).
3. Let Aut <- EvalUsageRuleSet(RS, Agent, Action, KI)
4. If Aut = "allowed", Execute (Action, KI, TGA).

Define EvalUsageRuleSet (RS, Agent, Action, KI}
5. Let RS <- Sort-by-Precedence(RS, decreasing).
6. Let A <- DefaultAuthorization.
7. For each R in RS,
   a) Let AR <- EvalRule(R, Agent, Action, KI);
   b) If AR = "Allowed" or AR = "Denied",
      return AR.
8. Return A.

**Figure 1 - Request Evaluation Algorithm**

To determine the final Authorization value, a conflict resolution strategy must be employed, which is in turn subject to Governance Rules.

This algorithm abstracts the essential decisions that must be made, to wit:

1. Who can formulate a rule for a given KI? – in line 1;
2. How are conflicts between rules resolved? – in line 5-8
3. What are the allowed actions over KIs? – in line 4.

We detail possible answers to these questions in the following sub-sections.

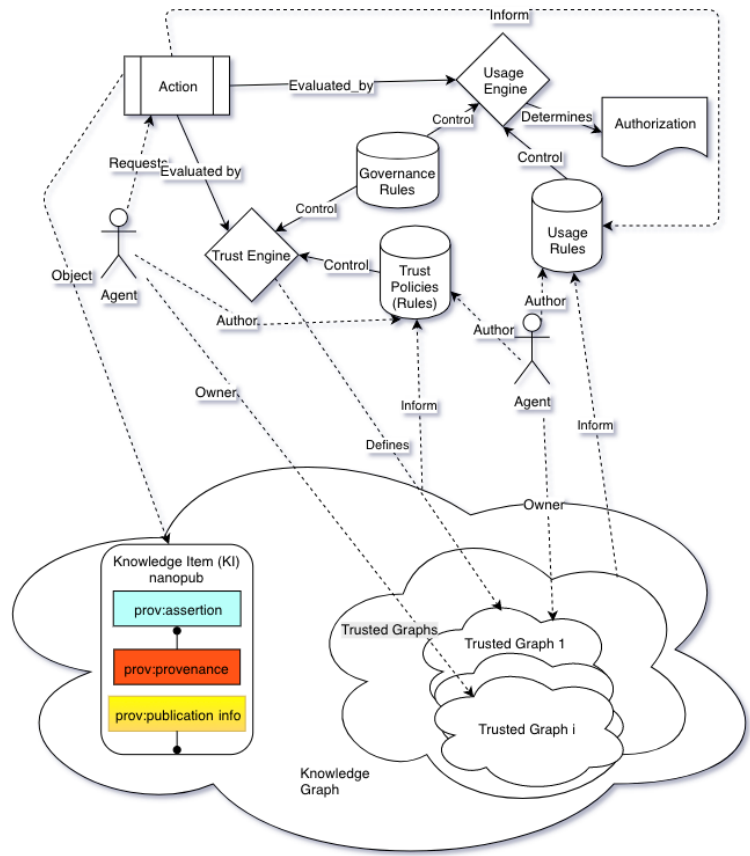

**Figure 2 Usage process for information in a KG**

## 4.3 Rules

The UsageRuleSet(KI) function determines what are the applicable rules given a KI. For trust rules, it is the author of the request. For privacy rules this corresponds to anwering the question "who has the right to define a Privacy Rule that controls actions over this KI?" The definition of Privacy itself indicates that it must be any agent that is somehow *related* to the information contained in the KI. Any useful instantiation of the framework must spell out what are the accepted relation types, which can include:

- An identification property for any agent that is included in the KI – for example, some person appearing in a posted photo or video;
- Any relation denoting referral to a person included in the KI – for example, some person cited in a post;
- Any creation or authorship relation;
- Any agent related to the creation of the KI – for example the author of a video posted by someone else;
- Any agent who has legal jurisdiction over an agent identified or mentioned in the KI;

The Agent representing the legal system(s) that has(have) jurisdiction over the KG, over the Agent or over the Action request.

The presence of such relations can directly occur in the KG (i.e., as a typed edge), or as a composition of valid relations. Furthermore, the rules may (or may not) allow the use of inferred relations in the KG.

Rules are of the form antecedent => consequent, both of which are sets of statements [1]. The antecedent of privacy rules may refer to any statements in the KG, including

- Any statements in the graphs in the KI's nanopublication (assertions, provenance and publication info);
- The identity of the agent requesting permission;
- The type of Action;
- Information in the KG serving as contextual information), such as
    - Date/time of the request;
    - Current location of agents involved;

Seen as nanopublications, the actual specification of the rule is given in its assertion graph, using a notation such as N3Logic **Error! Reference source not found.** or SWRL[7]. Governance Rules are Rules that include other Rules (in either of antecedent and consequent), so in this sense they are meta-rules.

## 4.4 Conflict resolution

The Sort-by-Precedence(RS, decreasing) function call sorts the enabled rules are sorted in descending precedence order, typically combining several types of information to establish order relations between rules. Some possible complementary order relations that can be employed are:

- Hierarchical relations between users – rules defined by a higher-ranked user take precedence over rules defined by lower-ranked ones. For example, rules established by laws take precedence over rules stated by individuals;
- Hierarchical relations between relation types. Rules defined by users related to the KI through a higher-ranked relation type take precedence over rules defined by users related via a lower-ranked relation type. For example, one may state that "identifies" takes precedence over "mentions". Thus, in a video where a person A is identified, and person B is mentioned in a conversation, person A's rules would take precedence over person B's rules.

Since hierarchies are partial orders, they may not completely define precedence, so further conflict resolution strategies are still needed. Such and Criado [8] identified six categories of strategies that can be employed. Most require user involvement at runtime, but the aggregation-based class can be easily incorporated into an algorithm. Strategies in this class define an aggregation function such as consensus, majority, minimum fixed number of votes, permit-overrides, deny-overrides, etc. – see also **Error! Reference source not found.** – and replace the set of conflicting rule results by a single aggregated result.

An alternative to the aggregation approach is to decompose the KI into finer-grained elements so that each is subject to only one rule. This makes sense when the Action to be performed can be stated as the composition of the same operation performed on each element independently. For the elements where the Action is denied, the resulting composite object is altered, such as blurring the face of a given person in a group photo.

## 5 Example Scenario revisited

In this section we re-examine the motivational scenario light of the KG usage framework, showing how the main important aspects are represented.

---

[7] https://www.w3.org/Submission/SWRL/

The first rule captures Ed's requirement about Non-Profits: he only trusts organizations that openly publish who are their financial officers, and their financial records. Financial records must be validated by accredited audit organizations.

The trust rule below captures this, expressed using N3Logic with extensions. We state under KG some statements we assume to be present in the KG:

**KG**

NonProfit subClassOf org:Organization. AuditCo subClassOf org:Organization. CertificationAgency subClassOf org:Organization. Donate subClassOf Action.
<ReliefOrg> type NonProfit. <AuditInc> type <CertificationAgency>.
<ReliefOrg> officer <George>. <AuditInc> officer <George>.

**RuleEd1**

{?O type org:Organization; hasOfficial: ?Ofc; hasFinancialRecord ?FR. ?Ofc type foaf:Person.
?FR assertions ?FRa. ?FRa log:semantics ?FRaS; ?FRAS log:includes
                { ?FR auditedBy ?Aud. ?Aud type AuditCo}.
?FR provenance ?PFr. ?PFr log:semantics ?PFrS. ?PFrS log:includes {prov:hasPrimarySouce ?DOCS}.
?DOCS assertions ?DOCSa. ?DOCSa log:semantics ?DOCSaS.
?DOCSaS log:includes {AuditCo certifiedBy ?CA. }. <TrustedGraphEd> author <Ed>
   log:semantics ?TGEd. ?TGEd log:includes {?CA type CertificationAgency}}
=> {<TrustedGraphEd> :add {?O type:NonProfit}. }

This is a trust rule since it refers to KIs which are needed as input for Ed to take an action, "Donate".
Ed's second rule is in fact a business rule – he does not want to contribute to an organization in which George is an officer. While this could be embedded in application code using the KG, we express it here as a rule as well, to facilitate discussion.

**RuleEd2**

{<TrustedGraphEd> author <Ed>; log:semantics ?TGEd. ?TGEd log:includes
    {?O type:NonProfit, officer:<George>}. <ruleEd1 author <Ed>}.
ruleEd1 assertions ?ARule1. ?Arule1 log:semantics ?Arule1S.?ARule1S log:includes
    {{ <act1> type Donate, recipient <?O>. <Ed> intends <act1>}
=> {<at> type Authorization, rule <RuleEd2>, action <act1>, value "Denied"}}

Next, we look at George's privacy rule. It is a privacy rule because it refers to an action over personal information about George.

**RuleGeorge1**

{G :is {?O type:NonProfit; officer:<George>. <TrustedGraphGeorge> author <George>; log:semantics
    ?TGG. ?TGG log:includes {?O officer:<George>. <ruleGeorge1> author <George>}.
ruleGeorge1 assertions ?AGeorge1. ?AGeorge1 log:semantics ?AGeorge1S. ?Ageorge1S log:includes
    { <act1> type Read; object ?G. ?A intends <act1>}
=> {<at> type Authorization; rule <RuleGeorge1>; action <act1>; value "Denied"}}

The stipulation expressed in the transparency legislation for non-profits to divulge its officers can be expressed as

**KG**

NonProfitAct type Law.

**RuleTransp**

{<ruleTransp> author <Congress>. <ruleTransp> assertions ?ATransp. ?ATransp log:semantics
    ?ATranspS. ?ATranspS log:includes{?G is {?O type Nonprofit; officer ?Ofr.
    <act1> type Read; object ?G. ?A intends <act1>}
=> {<at> type Authorization; rule <ruleTtransp>; action <act1>; value "Allowed"}}.
<ruleTransp> provenance ?PTransp. ?PTransp log:semantics ?PTranspS. ?PTranspS log:includes
    {<ruleTransp> prov:HasPrimarySource <NonProfitAct>}

We have also included in the provenance graph of RuleTransp a reference to the proper legal document, the text of the law (<NonProfitAct>) that is being interpreted by this rule.

In order to manage the conflict between RuleGeorge1 and RuleTransp, there is a meta-rule stipulating the latter has precedence over the former. This precedence relation is used in the Sort-by-Precedence function call in line 5 in Figure 1.

**KG**

PersonalPrivacyRule subClassOf Rule. Legislation subClassOf Rule.

**MetaRule1**

{<MetaRule1> assertions ?AMR1. ?AMR1 log:semantics ?AMR1S. ?AMR1S log:includes
    {{?R1 type PersonalPrivacyRule. ?R2 type Legislation} => {?R2 precedes ?R1}}

We can also define more precisely what is a PersonalPrivacyRule – it is any rule that uses a PersonalInformation property in its antecedent.

?RuleP assertions ?RulePA.?RulePA log:semantics ?RulePAS;?RulePAS log:includes {{?p1 ?r ?p2.
    (?p1 rdf:type Person OR ?p2 rdf:type Person)}}
=> {?r rdf:type PersonalInformationRelation}
{?RulePP assertions ?RulePAS. ?RulePPAS log:semantics ?RulePPASS;
?RulePPAS log:includes{?RulePP antecedent ?RulePPAA. ?RulePPAA log:semantics ?RulePPAAS
    ?RulePPAAS log:includes {{?p ?r ?q.(?r rdf:type PersonalInformationRelation)}}
=>{?RulePP type PersonalPrivacyRule}}

The last scenario does not require any additional rules. It simply results in an "allowed" authorization, because no conflict arises between RuleGeorge1 and RuleTransp when applied to <AuditInc>, since it is not of type NonProfit. This illustrates a possible loophole in the legislation which could be avoided if the NonProfit legislation prohibited an organization from being audited by another organization having common officers.

# 6   Conclusions

We have presented a usage framework that explicates the various types of specifications that must be made to capture privacy, transparency and trust concerns. The framework also provides a better understanding of the relations between these concerns.

Trust entails determining which data items will be used to perform an action; privacy and transparency involve controlling who can perform an action over a data item. Trust is thus more fundamental, as privacy and transparency rules must be based on trusted data. Furthermore, privacy and transparency are seen as being different points along the control dimension, thus explaining the natural tension between the two.

We have shown how legal requirements, and other types of norms, which ultimately regulate the functioning of any application that uses the KG, can be incorporated into the KG itself.

The various nuances and interdependencies of these concerns we illustrated in a running example. One interesting point in the example is the fact that in spite of careful policies, loopholes in the regulations could allow undesired actions to take place.

As ongoing and future work, we are investigating implementation architectures to allow efficient and scalable usage control over existing KGs, as well as exploring the applicability to various domains.

**ACKNOWLEDGMENTS**

Daniel Schwabe was partially supported by a grant from CNPq.Daniel Schwabe was partially supported by a grant from CNPq.